# SHAP Stability in Credit Risk Management: A Case Study in Credit Card Default Model

Luyun Lin , Yiqing Wang*

**Abstract:** The increasing development in the consumer credit card market brings substantial regulatory and risk management challenges. The advanced machine learning models' applications bring concerns about model transparency and fairness for both financial institutions and regulatory departments. In this study, we evaluate the consistency of one commonly used Explainable AI (XAI) technology, SHAP, for variable explanation in credit card probability of default models via a case study about credit card default prediction. The study shows the consistency is related to the variable importance level and hence provides practical recommendation for credit risk management.

## 1. Introduction

The consumer credit card market is a critical component of retail banking, offering unsecured credits to millions of consumers. In the United States along, nearly 4,000 issuers, together with dozens of co-brand merchant partners, provide credit card services to over 190 million consumers and the overall credit card debt has surpassed $1 trillion at the end of 2022[1].

While the vast market continues to generate significant profits and vitality for the financial industry, it also introduces considerable regulatory risks. The risk management in the credit market is highly data-driven and subject to strict regulatory oversight. Many advanced quantitative models, i.e., machine learning algorithms, and deep neural networks, are employed to estimate key risk parameters, such as Probability of Default (PD), which quantifies the likelihood whether customers will become delinquent or default in the future. Machine Learning (ML) models are often regarded as "black boxes" due to their complex, non-linear structures and lack of intuitive explanations, which poses challenges for compliance with regulatory standards that require clear justification of credit decisions[2], [3]. With the increasing adoption of advanced ML models, concerns regarding transparency and interpretability have drawn growing attention from regulatory authorities.

Recent regulatory guidelines, such as the Equal Credit Opportunity Act (ECOA)[4], the Fair Credit Reporting Act (FCRA)[5], and supervisory expectations from the Federal Reserve and the Consumer Financial Protection Bureau (CFPB), have increasingly highlighted the importance of model interpretability in credit decisioning. The Consumer Financial Protection Circular[6], for example, require creditors to provide statements of specific reasons to applicants against whom adverse action is taken. The clear and understandable reason statement for adverse credit actions could ensure fairness in algorithmic decision-making processes.

*Correspondence: woshilucy712@gmail.com

To address these challenges, Explainable Artificial Intelligence (XAI) techniques have become essential tools for enhancing model transparency while maintaining predictive performance. XAI enables financial institutions to interpret and communicate the internal logic of complex ML models, thereby supporting regulatory compliance, facilitating model governance, and ensuring fairness in credit decisioning[7], [8]. Among various XAI methods, SHAP[9] (Shapley Additive Explanations) has gained widespread adoption in credit risk modeling due to its ability to provide both global and local explanations. By attributing the prediction of a model to individual features in a consistent and additive manner, SHAP facilitates transparency and fairness in high-stakes financial decisions. However, despite its popularity, little attention has been paid to the stability of SHAP values, that is, the extent to which SHAP explanations remain consistent across different model runs. This aspect is crucial for regulatory validation, as unstable explanations can undermine the credibility of model interpretation and raise concerns about fairness and reliability.

This study aims to provide practical insights into the SHAP value stability for credit risk machine learning-based probability of default model. Using a real-world credit default dataset, we evaluate the stability performance of SHAP value and practical guidance about its applications.

## 2. Methodology

In this chapter, we introduce the experiment design and methodology used during this study as shown in the flowchart below. The end-to-end experiment design has three major sections inside: 1) Data Source and Data Cleaning; 2) Model Development; 3) XAI Evaluation.

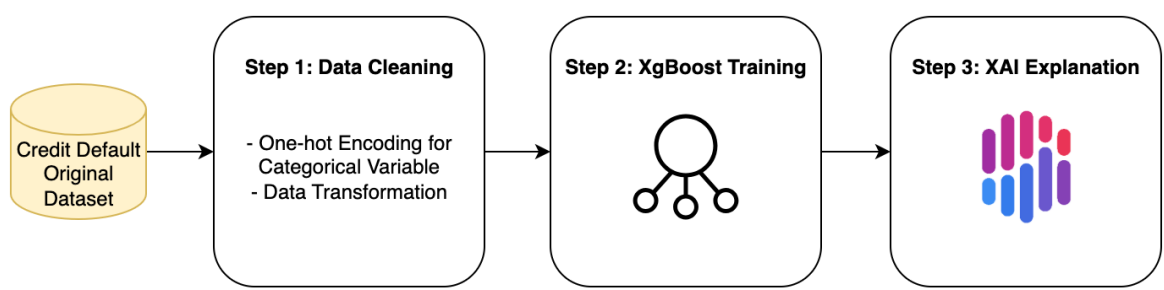

Figure 1: Experiment Design Flowchart

*2.1. Data Source and Cleaning*

We consider a credit card client's default dataset from UCI Machine Learning Repository[10]. This dataset contains unique 30,000 accounts' information in Taiwan from April 2005 to September 2005. Table 1 below lists all attributes and their metadata information. There are four major attribute categories included inside the dataset:1) account owner's demographic information; 2) account credit data; 3) account history of payments; 4) account history of bill statements. In total, 23 potential independent features are included, where 14 features are numerical values and the rest 9 features are categorical values with special meaning on each category.

Table 1: Variable Overview

| Variable Name | Description | Data Type | Data Range |
|---|---|---|---|
| id | Unique ID for Each Client | String | Not applicable |
| limit_bal | Amount of credit given in NT dollars (includes individual and family/supplementary credit | Numeric | 10,000 to 1,000,000 |
| sex | Gender (1=male, 2=female) | Category | 1 or 2 |
| education | Education Level(0 = others, 1=graduate school, 2=university, 3=high school, 4=others, 5=unknown, 6=unknown) | Category | 0,1,2,3,4,5,6 |
| marriage | Marital status (0=others, 1=married, 2=single, 3=others) | Category | 0,1,2,3 |
| age | Age in years | Numeric | 21 to 79 |
| pay_0 - pay_5 | Repayment status in each month from April to September 2005 (-2=No Consumption,-1=pay duly,0=The use of revolving credit 1=payment delay for one month, 2=payment delay for two months, … 8=payment delay for eight months, 9=payment delay for nine months and above) | Category | -2, -1, 0, 1, 2, 3, 4, 5, 6, 7, 8, 9 |
| bill_amt1 - bill_amt6 | Amount of bill statement in each month from April to September 2005 (NT dollar) | Numeric | from -339603 to 1664089 |
| pay_amt1 - pay_amt6 | The amount of previous payment in each month from April to September 2005 (NT dollar) | Numeric | from 0 to 1684259 |
| default.payment.next.month | Default payment (1=yes, 0=no) | Binary | 0, 1 |

The response variable is whether the account will have default payment in the next month (October 2005), where 1 means default and 0 means non-default. There are 23,364 accounts that are non-default while the rest 6,636 accounts are default for the next month's payment. It is noticed that the dataset is not well-balanced as only around 22.22% of accounts have defaulting value in the response variable. To prevent the effect of imbalance, over-sampling or down-sampling approaches, i.e., SMOTE[11], are used in the data processing process. However, the over-sampling approach may introduce inaccuracy to SHAP evaluation, while the down-sampling would adversely affect model performance. As the dataset is not an extremely imbalanced dataset, we do not use the over-sampling or down-sampling approach.

We perform Exploratory Data Analysis (EDA) and necessary data transformation on the original dataset. The data distribution on all numerical and categorical raw independent features are shown in Table 6 and Table 7 inside Appendix. Firstly, we combine several values in feature "Education" due to similarity. It has four classes (0,4,5,6) representing the similar group "other". Hence, we combine all of them together and the total percentage of combined group "other" is only 0.18%. Besides, it is noticed that the bill variables (bill_amt1-bill_amt6) have less than 2% accounts with negative values. It represents that the customer has overpaid or made early payments, resulting in a positive balance on their accounts. This may happen due to various reasons, i.e., overpayment, refunds, or early full payment. Hence, the data points are valuable to predict whether the customer will default in the next month, and we keep those observations as it is. Besides, we notice that the payment variable distribution (pay_1 – pay_6) is extremely right-skewed, which follows the consumer payment characters. Building a new payment variable by combining some extreme payment values (payment delay more than 3 months) may lead to better model performance. However, we keep the original classification as we want to explore whether the SHAP explanation will be largely affected by the variables with limited observation.

To ensure that all categorical variables are appropriately represented, we apply one-hot encoding to transform them into a binary format. This included variables such as sex, education level, marital status, and repayment status (PAY_1 to PAY_6). One-hot encoding is chosen to avoid introducing spurious ordinal relationships among categorical levels, thereby preserving the nominal nature of these features in the modeling process. In summary, the final dataset as input for the modeling process has 30,000 rows (representing 30,000 accounts) and 79 features.

*2.2. Model Developemnt*

The probability of default in credit risk modeling is a bivariate classification problem between good and bad accounts. Machine Learning technologies, i.e., SVM[12] and XgBoost[13], have shown a strong ability to capture non-linear relationships between independent variables and the default status. To amplify the model performance, we consider the XgBoost Model as the underlying algorithm for this study as it has better performance and faster speed compared with other ML models in credit risk area based on previous studies[14]. It is a supervised learning algorithm that combines decision tree models with the Gradient Boosting technique[15]. Gradient Boosting is an ensemble learning technique, which could build stronger predictive models by sequentially training weaker learners to minimize the pre-defined loss function.

To evaluate the XAI approach robustness, we train 100 different XgBoost models using the same pre-defined hyperparameters and various random seeds. We do not perform extensive hyperparameter tuning, as the study primary objective is not to optimize model performance but rather to evaluate the robustness and consistency of the XAI approach. The model's performance is strong with high discriminatory power and stable predictive metrics. Given the strong performance with default setting, further fine-tuning of hyperparameters would introduce unnecessary complexity without significantly improving the interpretability analysis. By avoiding overfitting to specific parameter choices, we ensure that our investigation into SHAP value stability more accurately reflects the intrinsic behavior of the model, rather than artifacts introduced by aggressive optimization.

After the model training, we compare the model's prediction with the exact response variable (1 or 0) on the test dataset using different metrics. Usually, in credit risk modeling area, we consider the following two major metrics: 1) Kolmogorov-Smirnov (KS) metric at 10 equally split bins based on volume; 2) confusion matrix at different thresholds and relevant metrics.

**Kolmogorov-Smirnov (KS)**

The Kolmogorov-Smirnov (KS) statistic is a widely used measure in credit risk modeling to assess the discriminatory power of binary classification models, i.e., the probability of default models[16]. It quantifies the maximum difference between the empirical cumulative distribution functions (empirical CDFs) of the positive (good) and negative (bad) classes. From a definition perspective, the formula is

$$KS = \max(TPR - FPR)$$

Where TPR is the cumulative true positive rate and FPR is the cumulative false positive rate.

In practice, especially for credit risk scorecards, the KS statistics are often calculated based on a decile-level table[17]. Accounts are sorted by prediction values (scores or probabilities) in ascending order and divided into ten equally sized bins regarding the volume. For each bin, the proportion of true good and bad accounts is calculated, followed by the cumulative distributions. The KS value is defined as the maximum absolute difference between the two cumulative distributions among all bins.

A key advantage of the decile-level KS approach is that it does not require a predefined classification threshold to calculate TPR and FPR as shown above. Instead, it evaluates the model effectiveness across the entire score range based on the relative ranking of scores. This makes it more robust and flexible, especially when the optimal cut-off point is unknown or varies based on the following usage. A higher KS value indicates better separation, and therefore better discriminatory power. Usually, KS between 0.3 to 0.7 is typically considered acceptable in credit risk modeling; values below 0.2 may indicate poor model performance, while values above 0.8 could suggest potential overfitting.

**Confusion Matrix and Corresponding Metrics**

The performance for binary classification models is commonly evaluated using a confusion matrix, a 2*2 tabular summary contributed by Kuhn and Johnson[18]. It compares the model's predicted labels with the actual ground truth given a pre-determined bad/good threshold. The confusion matrix consists of four major entries: true positives (TP), false positives (FP), true negatives (TN), and false negatives (FN). These components serve as the foundation for several important performance metrics commonly used in model evaluation as shown below.

Table 2: Confusion Matrix Illustration

|  |  | Predictions | | |
|---|---|---|---|---|
|  |  | **Positive (Bad/event)** | **Negative (Not Bad/ non-event)** |  |
| **Actuals** | **Positive (Bad/event)** | True Positive (TP) | False Negative (FN) | Recall (True Positive Rate) = TP/(TP+FN) |
|  | **Negative (Not Bad/ non-event)** | False Positive (FP) | True Negative (TN) | False Positive Rate = FP/(FP + TN) |
|  |  | Precision = TP/(TP+FP) | Negative predictive value = TN/(TN+FN) | F-score = 2*Recall*Precision/(Recall + Precision) |

In loan default prediction, evaluating model performance requires careful consideration of classification metrics that reflect both business impact and statistical accuracy[19]. Among these, precision, recall, and the F1-score are particularly relevant. Precision is defined as the proportion of true positive predictions among all instances predicted as defaulters. In this setting, high precision implies that the model minimizes the risk of misclassifying creditworthy borrowers as defaulters, thereby reducing opportunity loss due to unnecessary loan rejections. Recall, or sensitivity, measures the proportion of actual defaulters that are correctly identified by the model. High recall is critical for capturing as many truly risky borrowers as possible, thus helping lenders reduce potential financial loss from defaults. However, precision and recall often exhibit a trade-off, especially in imbalanced dataset like the one this paper is working on. To balance this trade-off, the F1-score is employed, which is the harmonic means of precision and recall. It provides a single metric that reflects the model's ability to accurately and comprehensively identify defaulters, making it particularly suitable when both types of misclassifications—false positives (rejecting good borrowers) and false negatives (approving risky ones)—incur non-negligible costs. As the F1-score offers a balanced and interpretable measure for model performance, we consider it the major metric when evaluating models.

Another crucial evaluation metric this paper considers is the Area Under the Receiver Operating Characteristic Curve (AUROC). The ROC curve plots the true positive rate against the false positive rate at various threshold settings. The AUROC score, ranging from 0 to 1, represents the probability that a randomly chosen positive instance is ranked higher by the model than a randomly chosen negative one. An AUROC of 0.5 indicates no discriminative power (equivalent to random guessing), whereas a value closer to 1 indicates excellent separability between classes. Unlike threshold-dependent metrics such as F1-score, AUROC provides a threshold-independent evaluation of model discrimination.

*2.3. Explainable Artificial Intelligence (XAI) Evaluation*

Complex Machine Learning Models, i.e., XgBoost and Deep Neural Network, are usually considered as "black boxes" because the internal decision-making processes are not as easily interpretable as easy as those of simpler models like logistic or linear regression. While these models can achieve relatively higher predictive accuracy, the lack of transparency raises concerns for model users. In credit decision-making, adverse actions—such as loan denials or unfavorable lending terms—require lenders to provide clear justifications under regulations like the Fair Credit Reporting Act (FCRA) and the Equal Credit Opportunity Act (ECOA). As more complex machine learning models are utilized, the lack of transparency poses challenges in meeting the regulatory requirement. Hence, XAI plays a critical role by providing interpretable insights into how models arrive at the denial or approval decisions. It enables institutions to identify key drivers behind adverse outcomes, ensure decisions are fair and free from bias, and enable them to communicate with consumers in a compliant manner.

SHAP[20] (SHapley Additive exPlanations) is a widely used XAI method that provides both global and local interpretability of complex machine learning models with formula below. It is inspired by the Shapley values from cooperative game theory, which offers a fair allocation of each feature's contribution to a model's prediction. By quantifying the impact of individual features on specific predictions, SHAP enables a clear understanding of model behavior at the instance level. This local interpretability makes SHAP particularly valuable in credit risk modeling, where it is important to justify individual lending decisions. For example, SHAP can explain why a loan application was rejected by identifying key risk factors—such as a low credit score—that had the most influence on the model's output.

Besides XAI transparency, the robustness of XAI methods is also critical for ensuring consistent, reliable, and legally defensible model explanations[21]. As required by the regulatory institutions, clear and consistent reasons for adverse action should be provided. If XAI explanations vary significantly across model recalibrations process, it can undermine true and lead to compliance risks. Robust XAI methods ensure that explanations remain stable across model perturbations, enabling fair and repeatable assessments of risk factors. This consistency is vital for producing adverse action notices, validating model fairness, and satisfying internal and external audits. Therefore, robustness in XAI is not just a technical requirement but a foundational aspect of trustworthy and responsible credit risk management.

To evaluate SHAP robustness at the global level, we compute each feature's SHAP value and corresponding rank across 100 independently trained models initialized with different random seeds. To quantify the SHAP robustness, we consider Kendall's W (Coefficient of Concordance) quantifies the agreement among multiple raters (rankers) with formula below[22].

$$W = \frac{12 * S}{m^2 * (n^3 - n)}$$

where
- n is the number of features being ranked
- m is the number of ranking lists, which is the number of models in this study
- $S = \sum_{i=1}^{n}(R_i - \bar{R})^2$ where $R_i$ is the sum of ranks for feature $i$ across all models and $\bar{R}$ is the average of all $R_i$

The value of the Kendall's W ranges from 0 to 1, where higher value means stronger consistency.

## 3. Results

*3.1 Performance Evaluation*

For credit default binary classification, KS metric is one of the most used metrics in industry. Higher KS represents a stronger ability to distinguish bad and good customers. The histogram below shows the KS value based on 10-bins decile dataset among the 100 trained models. The absolute KS value is relatively high, showing the strong prediction ability. Besides, the KS range is pretty narrow, ranging from 41.31% to 43.63%, with median value at 42.37%. It indicates the trained models with different random seeds have similar prediction performance.

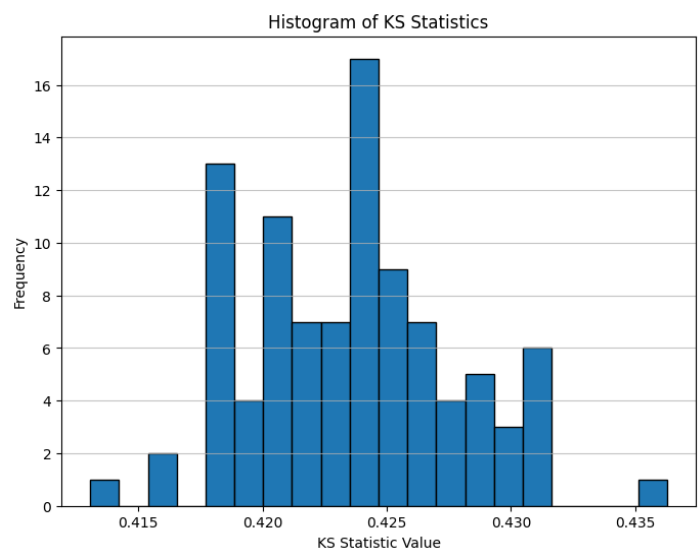

Figure 2: KS Performance

The figure below shows the distribution of metrics among the 100 trained XgBoost models and table below shows the confidence interval and median value for each metric. It is noticed that the model performance regarding those confusion matrix related metrics, like accuracy, sensitivity, F-1 Score, and AUROC are constant among the 100 trained models. The median accuracy is around 79.5%, showing strong powerful performance for the model. Considering there is no sophisticated hyperparameter tuning process, the model still performs similarly with another benchmark analysis[23].

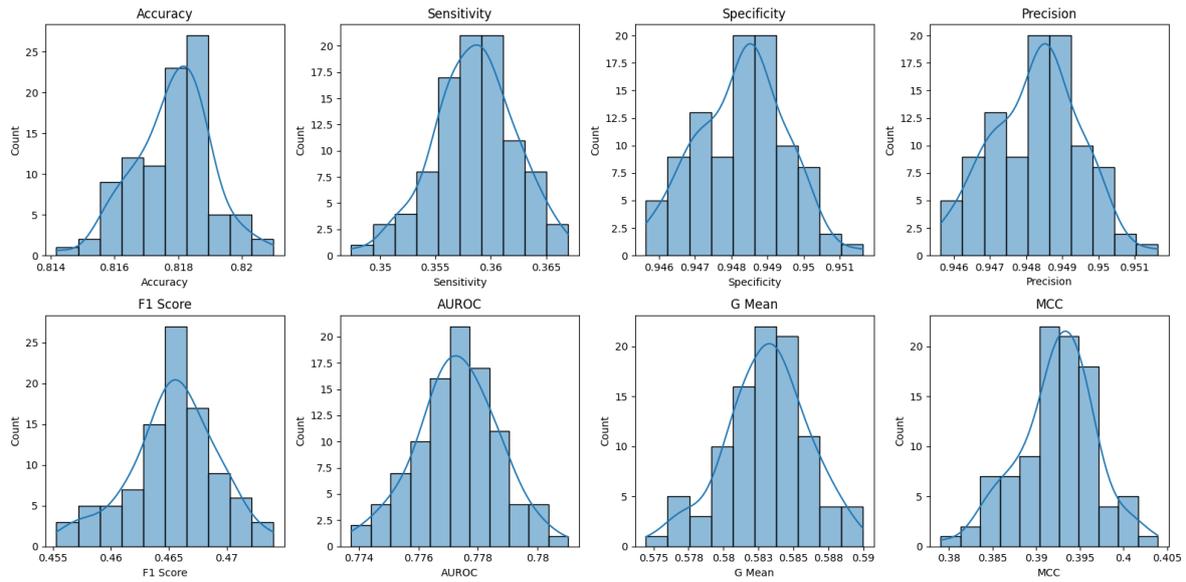

Figure 3: Confusion Metric Distribution Plots

Table 3: Confusion Metric Median and 95% Confidence Interval (CI)

| Metric | Median | Lower 95% CI | Upper 95% CI |
| --- | --- | --- | --- |
| Accuracy | 79.53% | 79.25% | 79.77% |
| Sensitivity | 53.73% | 53.08% | 54.26% |
| Specificity | 86.86% | 86.68% | 87.01% |
| Precision | 86.86% | 86.68% | 87.01% |
| F1 Score | 53.73% | 53.08% | 54.26% |
| AUROC | 77.73% | 77.46% | 78.00% |
| G Mean | 68.32% | 67.83% | 68.71% |
| MCC | 40.59% | 39.76% | 41.27% |

Plot below shows the ROC curve among the 100 trained models. As specified above, the AUROC value ranges from 0.774 to 0.781, showing strong ability for positive and negative class discrimination. Similarly, the ROC curve performs better than the existing benchmark analysis.

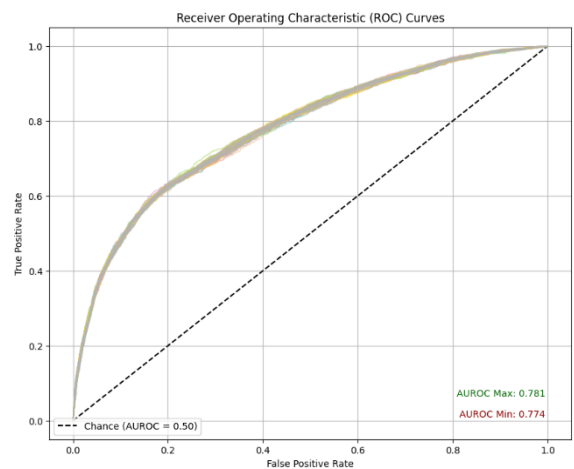

Figure 4: ROC Curve

*3.2 Gloal SHAP Evaluation*

Plot below is the heatmap of SHAP value among the 100 trained models among all independent features. It is noticed that the exact SHAP value uses positive or negative signs to represent the variable's contribution direction to the final output. In our study, for example, positive SHAP values means the features contribute to the predicted probability of default, while negative values represent the negative relationship with the predicted probability of default. To evaluate the global SHAP value, we calculate it as the sum of absolute SHAP value among all data points for each model. Then, we calculate the rank of each global SHAP value among all the 100 models for each feature. The lower the rank, the higher the global SHAP value is, representing the higher importance level.

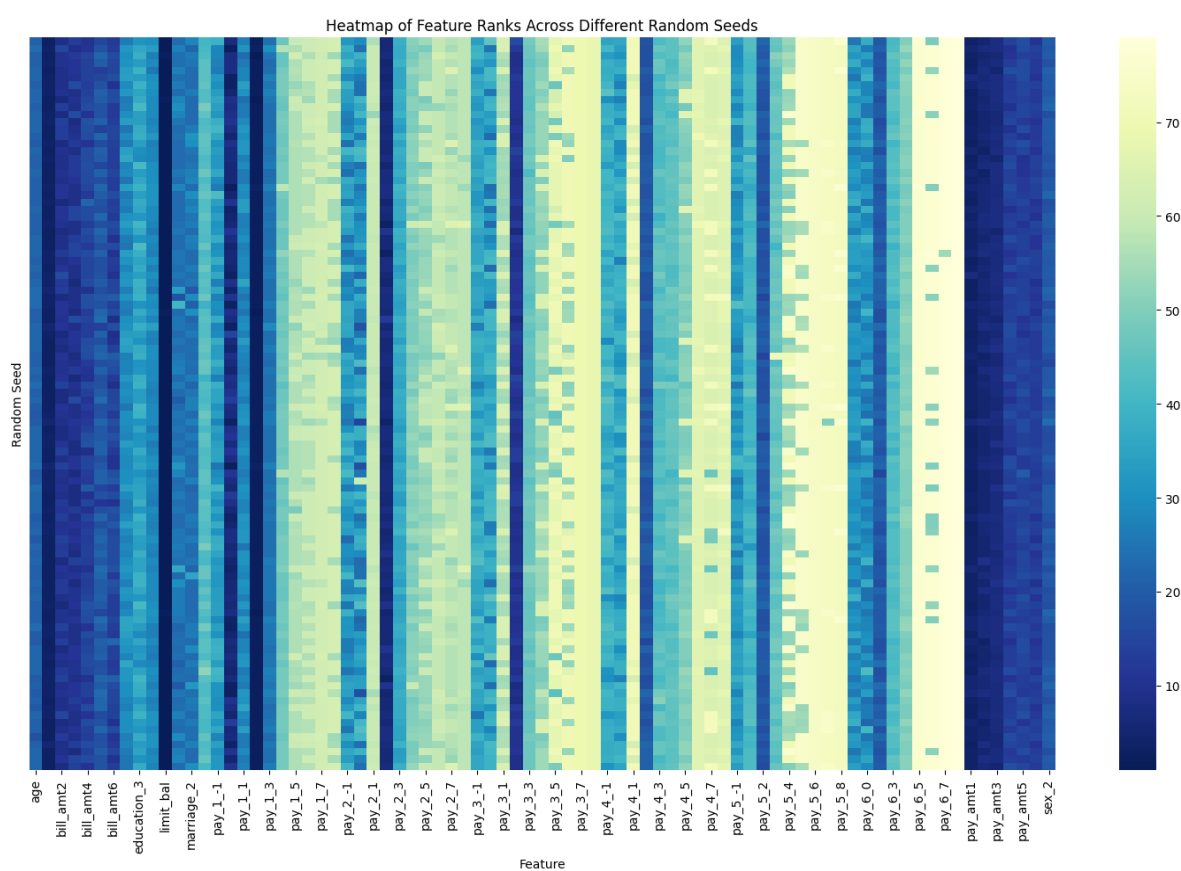

Figure 5: Heatmap of SHAP rank for each variable among models

We notice the SHAP rank is consistent among models for some variables. Tables below show the rank value counts for the first five variables with the highest mean SHAP ranks and that for the first five variables with the lowest mean SHAP ranks. For example, the variable limit_bal is the amount of given credit in NT dollars. Out of 100 SHAP ranks, the variable has 96 ranking at 1 and 4 at 2. It shows this variable has stable and significant SHAP value among all models. As shown below, those variable's rank is stable among different models. The extreme variable, pay_6_8 (indicator showing the account has payment delay for 8 months in April 2025), always has the lowest rank time among all models.

Table 4: List of five variables with highest SHAP mean rank

| Variable | Meaning | Rank | Frequence |
|---|---|---|---|
| Limit_bal | Amount of credit given in NT dollars (includes individual and family/supplementary credit | 1 | 96 |
| | | 2 | 4 |
| pay_1_2 | Repayment status in September 2005: Payment delay for two months | 1 | 4 |
| | | 2 | 93 |
| | | 3 | 3 |
| bill_amt1 | Amount of bill statement in September, 2005 (NT dollar) | 3 | 56 |
| | | 4 | 39 |
| | | 5 | 5 |
| pay_amt1 | Amount of previous payment in September, 2005 (NT dollar) | 3 | 29 |
| | | 4 | 49 |
| | | 5 | 17 |
| | | 6 | 3 |
| | | 7 | 2 |
| pay_amt2 | Amount of previous payment in August, 2005 (NT dollar) | 4 | 5 |
| | | 5 | 46 |
| | | 6 | 25 |
| | | 7 | 16 |
| | | 8 | 7 |
| | | 9 | 1 |

Table 5: List of five variables with the lowest SHAP mean rank

| Variable | Meaning | Rank | Frequence |
|---|---|---|---|
| pay_5_7 | Repayment status in May 2005: Payment delay for seven months | 50 | 1 |
| | | 73 | 85 |
| | | 75 | 14 |
| pay_5_6 | Repayment status in May 2005: Payment delay for six months | 73 | 1 |
| | | 74 | 99 |
| pay_6_5 | Repayment status in April 2005: Payment delay for five months | 76 | 81 |
| | | 77 | 19 |
| pay_6_7 | Repayment status in April 2005: Payment delay for seven months | 54 | 1 |
| | | 78 | 99 |
| pay_6_8 | Repayment status in April 2005: Payment delay for eight months | 79 | 100 |

On the other hand, for variables with relatively middle SHAP rank among all features, the SHAP rank is not very stable among different models. The graph below shows the unique SHAP rank and distributions for the six variables with the highest unique rank frequency among the 100 models. Their average rank ranges from 24 to 37. We notice the extreme case, pay_2_0 (The repayment status in August 2025 is the use of revolving credit), has 25 different ranks among the 100 models, with lowest rank at 61 and highest rank at 15.

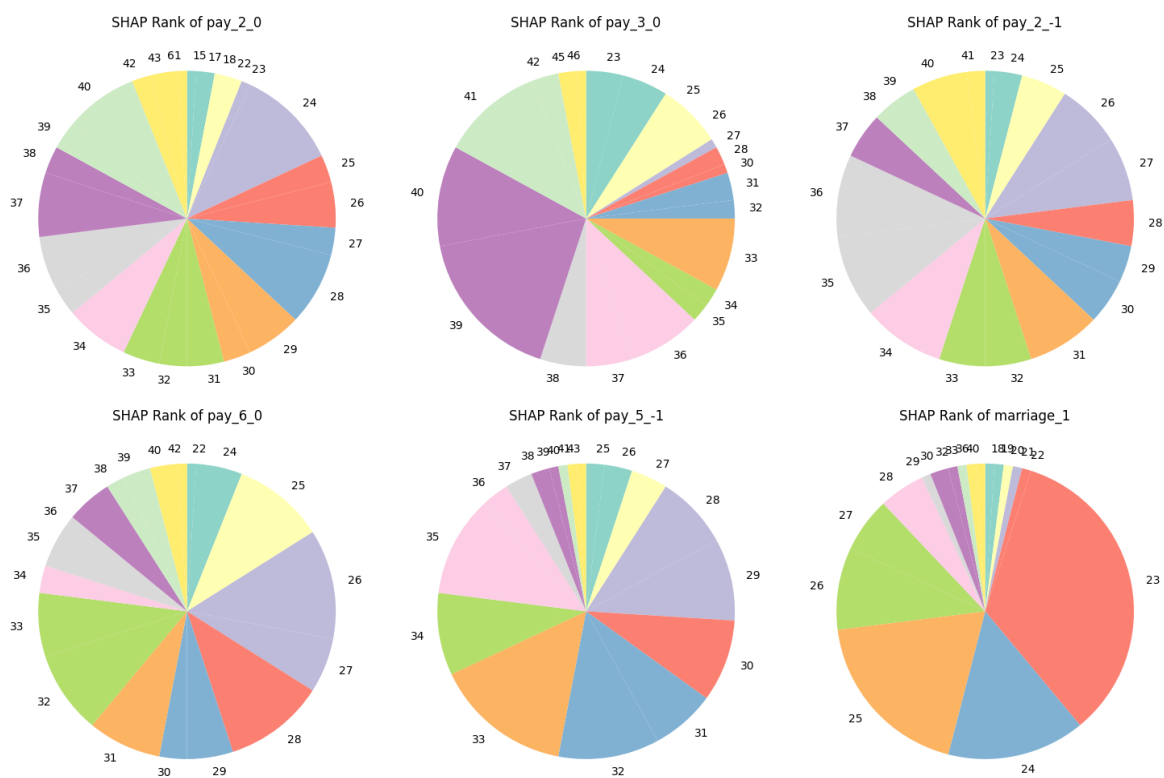

Figure 6: Unique SHAP Rank Distribution for Inconsistent Variables

Besides, we notice that Kendall's W statistic value for all features is 0.9775 and the chi-square test p-value is less than 0.001. It represents with significance level of 0.05, the probability that the ranking is random is less than 0.001, which means the ranking is much more consistent than a random ranking. For subgroup features, like pay status indicator, bill amount, or pay amount, the Kendall's W value is also relatively high, and the corresponding chi-square test p-value is less than 0.001. It shows the SHAP ranking is consistent among all trained models.

## 4. Conclusion

In this study, we investigate the stability of SHAP approach on feature explanation in credit risk modeling during a case study using credit card default dataset in Taiwan. The results show that variables with the strongest or lowest prediction power tend to have stable SHAP rankings among different models. Conversely, variables with middle contributions often exhibit high variability in their SHAP-based importance rankings. The Kendall's W statistics also suggest that the SHAP ranking is not purely stochastic but instead reflects the underlying predictive strength of each variable.

From a regulatory standpoint, these findings carry important implications for model explainability, particularly in the context of adverse action notices. If a variable appears among the top-ranked features by SHAP but its ranking is highly unstable, it may not be appropriate to rely on it as a consistent driver of model decisions. This highlights the need for additional scrutiny before using such features in customer-facing explanations or regulatory reporting.

This study has certain limitations. First, it is only based on a single dataset and a single model class (XgBoost) without hyperparameter training, which may limit the generalizability of the conclusions. Future research should explore SHAP stability across multiple datasets and more machine learning types. Such broader analysis would support more robust and reliable guidance for adverse action explanations and help align machine learning practices with regulatory expectations.

By revealing the relationship between variable strength and SHAP stability, this study contributes to the growing body of research on explainable AI in credit risk and underscores the importance of evaluating explanation reliability alongside traditional model performance metrics.

**Conflicts of Interest:** Authors confirm that there is no conflict of interest with anyone involved.

**Supplementary Materials**

Table 6: Continuous Raw Variable Descriptive Statistics

|  | count | mean | std | min | 25% quantiles | Median | 75% quantiles | max |
|---|---|---|---|---|---|---|---|---|
| limit_bal | 30000 | 167484.32 | 129747.66 | 10000 | 50000 | 140000 | 240000 | 1000000 |
| age | 30000 | 35.48 | 9.21 | 21 | 28 | 34 | 41 | 79 |
| bill_amt1 | 30000 | 51223.33 | 73635.86 | -165580 | 3558.75 | 22381.5 | 67091 | 964511 |
| bill_amt2 | 30000 | 49179.07 | 71173.76 | -69777 | 2984.75 | 21200 | 64006.25 | 983931 |
| bill_amt3 | 30000 | 47013.15 | 69349.38 | -157264 | 2666.25 | 20088.5 | 60164.75 | 1664089 |
| bill_amt4 | 30000 | 43262.94 | 64332.85 | -170000 | 2326.75 | 19052 | 54506 | 891586 |
| bill_amt5 | 30000 | 40311.40 | 60797.15 | -81334 | 1763 | 18104.5 | 50190.5 | 927171 |
| bill_amt6 | 30000 | 38871.76 | 59554.10 | -339603 | 1256 | 17071 | 49198.25 | 961664 |
| pay_amt1 | 30000 | 5663.58 | 16563.28 | 0 | 1000 | 2100 | 5006 | 873552 |
| pay_amt2 | 30000 | 5921.16 | 23040.87 | 0 | 833 | 2009 | 5000 | 1684259 |
| pay_amt3 | 30000 | 5225.68 | 17606.96 | 0 | 390 | 1800 | 4505 | 896040 |
| pay_amt4 | 30000 | 4826.07 | 15666.15 | 0 | 296 | 1500 | 4013.25 | 621000 |
| pay_amt5 | 30000 | 4799.38 | 15278.30 | 0 | 252.5 | 1500 | 4031.5 | 426529 |

| pay_amt6 | 30000 | 5215.50 | 17777.46 | 0 | 117.75 | 1500 | 4000 | 528666 |

Table 7: Categorical Raw Variable Descriptive Statistics

| Variable | Number Value | Meaning | Count | Percentage |
|---|---|---|---|---|
| **Sex** | 1 | Male | 11888 | 39.63% |
| | 2 | Female | 18112 | 60.37% |
| **Education** | 0 | Others | 14 | 0.05% |
| | 1 | Graduate School | 10585 | 35.28% |
| | 2 | University | 14030 | 46.77% |
| | 3 | High School | 4917 | 16.39% |
| | 4 | Others | 123 | 0.41% |
| | 5 | Unknown | 280 | 0.93% |
| | 6 | Unknown | 51 | 0.17% |
| **Marriage** | 0 | Others | 54 | 0.18% |
| | 1 | Married | 13659 | 45.53% |
| | 2 | Single | 15964 | 53.21% |
| | 3 | Divorce | 323 | 1.08% |
| **Pay_1** | -2 | No Consumption | 2759 | 9.20% |
| **(Repayment status in September, 2005)** | -1 | Paid in full | 5686 | 18.95% |
| | 0 | The use of revolving credit | 14737 | 49.12% |
| | 1 | Payment delay for 1 month | 3688 | 12.29% |
| | 2 | Payment delay for 2 month | 2667 | 8.89% |
| | 3 | Payment delay for 3 month | 322 | 1.07% |
| | 4 | Payment delay for 4 month | 76 | 0.25% |
| | 5 | Payment delay for 5 month | 26 | 0.09% |
| | 6 | Payment delay for 6 month | 11 | 0.04% |
| | 7 | Payment delay for 7 month | 9 | 0.03% |
| | 8 | Payment delay for 8 month | 19 | 0.06% |
| **Pay_2** | -2 | No Consumption | 3782 | 12.61% |
| **(Repayment status in August, 2005)** | -1 | Paid in full | 6050 | 20.17% |
| | 0 | The use of revolving credit | 15730 | 52.43% |
| | 1 | Payment delay for 1 month | 28 | 0.09% |
| | 2 | Payment delay for 2 month | 3927 | 13.09% |
| | 3 | Payment delay for 3 month | 326 | 1.09% |
| | 4 | Payment delay for 4 month | 99 | 0.33% |
| | 5 | Payment delay for 5 month | 25 | 0.08% |
| | 6 | Payment delay for 6 month | 12 | 0.04% |
| | 7 | Payment delay for 7 month | 20 | 0.07% |
| | 8 | Payment delay for 8 month | 1 | 0.00% |
| **Pay_3** | -2 | No Consumption | 4085 | 13.62% |
| **(Repayment status in July, 2005)** | -1 | Paid in full | 5938 | 19.79% |
| | | 0 | The use of revolving credit | 15764 | 52.55% |
| | | 1 | Payment delay for 1 month | 4 | 0.01% |
| | | 2 | Payment delay for 2 month | 3819 | 12.73% |
| | | 3 | Payment delay for 3 month | 240 | 0.80% |
| | | 4 | Payment delay for 4 month | 76 | 0.25% |
| | | 5 | Payment delay for 5 month | 21 | 0.07% |
| | | 6 | Payment delay for 6 month | 23 | 0.08% |
| | | 7 | Payment delay for 7 month | 27 | 0.09% |
| | | 8 | Payment delay for 8 month | 3 | 0.01% |
| **Pay_4** | -2 | No Consumption | 4085 | 13.62% |
| **(Repayment status in June, 2005)** | -1 | Paid in full | 5938 | 19.79% |
| | 0 | The use of revolving credit | 15764 | 52.55% |
| | 1 | Payment delay for 1 month | 4 | 0.01% |
| | 2 | Payment delay for 2 month | 3819 | 12.73% |
| | 3 | Payment delay for 3 month | 240 | 0.80% |
| | 4 | Payment delay for 4 month | 76 | 0.25% |
| | 5 | Payment delay for 5 month | 21 | 0.07% |
| | 6 | Payment delay for 6 month | 23 | 0.08% |
| | 7 | Payment delay for 7 month | 27 | 0.09% |
| | 8 | Payment delay for 8 month | 3 | 0.01% |
| **Pay_5** | -2 | No Consumption | 4546 | 15.15% |
| **(Repayment status in May, 2005)** | -1 | Paid in full | 5539 | 18.46% |
| | 0 | The use of revolving credit | 16947 | 56.49% |
| | 2 | Payment delay for 2 month | 2626 | 8.75% |
| | 3 | Payment delay for 3 month | 178 | 0.59% |
| | 4 | Payment delay for 4 month | 84 | 0.28% |
| | 5 | Payment delay for 5 month | 17 | 0.06% |
| | 6 | Payment delay for 6 month | 4 | 0.01% |
| | 7 | Payment delay for 7 month | 58 | 0.19% |
| | 8 | Payment delay for 8 month | 1 | 0.00% |
| **Pay_6** | -2 | No Consumption | 4895 | 16.32% |
| **(Repayment status in April, 2005)** | -1 | Paid in full | 5740 | 19.13% |
| | 0 | The use of revolving credit | 16286 | 54.29% |
| | 2 | Payment delay for 2 month | 2766 | 9.22% |
| | 3 | Payment delay for 3 month | 184 | 0.61% |
| | 4 | Payment delay for 4 month | 49 | 0.16% |
| | 5 | Payment delay for 5 month | 13 | 0.04% |
| | 6 | Payment delay for 6 month | 19 | 0.06% |
| | 7 | Payment delay for 7 month | 46 | 0.15% |
| | 8 | Payment delay for 8 month | 2 | 0.01% |
| **default.payment.next.month** | 0 | Non-Default | 23364 | 77.88% |
| | 1 | Default | 6636 | 22.12% |